\documentstyle[pra,aps,graphicx,twocolumn]{revtex}
\begin{document}
\draft
\title{Coherent acceleration of Bose-Einstein condensates}
\author{Sierk P\"otting$^{1,2,3}$
        \thanks{email: sierk.poetting@wotan.opt-sci.arizona.edu},
        Marcus Cramer$^{1,4}$,
        Christian H. Schwalb$^{1,4}$,
        Han Pu$^1$,
        and Pierre Meystre$^1$}
\address{$^1$Optical Sciences Center, University of Arizona, Tucson, Arizona
85721}
\address{$^2$Max--Planck--Institut f\"ur Quantenoptik, 85748 Garching,
Germany}
\address{$^3$Sektion Physik, Universit\"at M\"unchen, 80333 M\"unchen,
Germany}
\address{$^4$Fachbereich Physik der Philipps--Universit\"at, Marburg,
Germany}

\date{\today}

\maketitle

\begin{abstract}
We present a theoretical analysis of the coherent acceleration of
atomic Bose-Einstein condensates. A first scheme relies on the
`conveyor belt' provided by a frequency-chirped optical lattice.
For potentials shallow enough that the condensate is not
fragmented, the acceleration can be interpreted in terms of
sequential Bragg scattering, with the atomic sample undergoing
transitions to a succession of discrete momentum sidemodes. The
narrow momentum width of these sidemodes offers the possibility to
accelerate an ultracold atomic sample such as e.g. a Bose-Einstein
condensate without change in its momentum distribution. This is in
contrast to classical point particles, for which this kind of
acceleration leads to a substantial heating of the sample. A
second scheme is based on the idea of a synchronous particle
accelerator consisting of a spatial array of quadrupole traps.
Pulsing the trapping potential creates a traveling trap that
confines and accelerates the atomic system. We study this process
using the concept of phase stability.
\end{abstract}

 \narrowtext

%===========================================
\section{Introduction}
\label{sec:Intro}
%===========================================

A number of applications of atom optics, from inertial sensors
\cite{Gust97} to nanofabrication, atomic holography
\cite{FujMitShi00,Sor97,Zoba99} and certain schemes for quantum
computation \cite{Jaks99,Bren99}, can benefit substantially from
manipulating quantum degenerate atomic beams with finely
controlled potentials. In particular, the atomic de Broglie
wavelength can be easily modified by accelerating or decelerating
an atomic beam to the appropriate velocity, thereby making it a
tunable source. There is also an increased need to control
ultracold samples of atoms in confined geometries, as e.g. in
atomic waveguides for potential integrated atom optics
applications. First transport experiments in waveguides were
recently carried out with non--condensed atomic samples, both in
stationary and in time--dependent potentials
\cite{Mull99,Key00,Dekk00,Hans01}.

The simplest and most important coherence property of an atomic
beam is its monochromaticity. With this in mind, our goal in this
paper is to determine the extent to which it is possible to
significantly accelerate an initially quasi-monochromatic atomic
sample without losing this property. For this purpose it is
sufficient to describe the atomic system in the Hartree mean-field
limit, remembering that this ``classical-like'' description makes
implicit assumptions on the quantum statistical properties of the
Schr{\"o}dinger field, essentially assuming that it is in a
coherent state. Other, more subtle aspects of the quantum
coherence of accelerated matter-wave fields
\cite{Gold98a,Gold98b,Prat99,Mey01} will be addressed in future
work.

We consider two different acceleration schemes: in the first one,
an accelerated trap is provided by a frequency-chirped optical
lattice, while in the second one we consider a synchronous
accelerator consisting of a spatial array of quadrupole traps,
which is similar to a scheme that was successful in manipulating the
velocity of neutral molecules and atoms \cite{Beth00,Madd99}. We
restrict our discussion to one dimension for simplicity.

Ultracold atoms in both static and time-dependent optical lattices
have been investigated in several contexts in the recent past. At
the single-atom level, they were exploited extensively in
theoretical and experimental work aiming at demonstrating effects
such as Bloch oscillations \cite{Daha96,Peik97a,Peik97b}, Landau-Zener
tunneling \cite{Niu96,Bhar97}, the appearance of Wannier-Stark
ladders \cite{Niu96,Wilk96,Madi97,Zapa01}, quantum chaos
\cite{Oska99} and the dynamics of mesoscopic quantum
superpositions \cite{Hayc00}. The extension of this work to
Bose-Einstein condensates \cite{Berg98,Choi99} includes the first
demonstration of a mode-locked atom laser \cite{Ande98}, a device
that can also be interpreted in terms of Landau-Zener tunneling.
Bragg scattering of Bose-Einstein condensates off a periodic
optical potential was also used in experiments to characterize
their coherence and phase properties \cite{Sten99,Hagl99}.  It was
exploited to generate multiple momentum sidemodes of the
condensate used in matter-wave four-wave mixing experiments
\cite{Deng99,Gold99,Trip00} and to realize atomic beam splitters
used in coherent matter-wave amplifiers \cite{Inou99,Kozu99}.

Of particular relevance in the present context is Ref.
\cite{Peik97b}, which explicitly considers the acceleration of
atoms in a moving periodic potential and interprets it in terms of
momentum transfer via multiple adiabatic rapid passage. This
reference also reports experiments demonstrating the acceleration
of atomic molasses with large initial momentum spreads as well as 
of ensembles with subrecoil momentum distributions. The final momentum spread
is limited by the the lattice recoil energy for the broad samples and
by the initial distribution for the subrecoil atoms, see Figs. 11 
and 13 of Ref. \cite{Peik97b}. We will show that we can suppress heating 
even in the case the atomic temperature is substantially lower than
the recoil temperature associated with the lattice, which in
particular holds for a Bose-Einstein
condensate and its quasi-monochromatic momentum distribution.
\footnote{Shortly after submission of this work, this
prediction was independently confirmed in experiments on the
acceleration of a condensate by a frequency-chirped optical
lattice \cite{Mors01}.}

The paper is organized as follows: Section \ref{sec:OpLat}
investigates the acceleration of an ultracold atomic ensemble in a
frequency-chirped optical lattice. We set the stage by first
considering an ensemble of {\it classical} point particles spread
over a large number of lattice periods. We compare its dynamics to
that of a Bose-Einstein condensate of the same spatial extent,
described in the Hartree mean-field regime. We demonstrate that
while the accelerated classical particles undergo considerable
heating, which is however substantially eliminated by the quantum
interferences associated with the extended atomic wave functions
of ultracold samples. The quantum dynamics are interpreted as
continuous Bragg scattering in a lattice with time--dependent
detuning. We briefly review the formalism of Bragg scattering
within the framework of a partitioned momentum space {\em et al.}
\cite{Blak00}, which allows one to discuss the acceleration in
terms of a coupled-mode approach that leads to a determination of
optimum acceleration parameters. We conclude this section by some
comments on the role of the mean-field interaction. Section
\ref{sec:Sync} then turns to the synchronous acceleration scheme.
Again, we first consider the acceleration of non-interacting
classical point particles, this time using the concept of phase
stability. We then study numerically the acceleration of a
condensate. Finally, section \ref{sec:SumCon} is a conclusion and
outlook.

%===========================================
\section{Accelerated optical lattice}
\label{sec:OpLat}
%===========================================

%-------------------------------------------
\subsection{Classical acceleration}
\label{subsec:classical}
%-------------------------------------------

To set the stage for our discussion, we first consider the
acceleration of an ensemble of non-interacting, point-like
classical particles in a frequency-chirped optical lattice formed
by two counterpropagating laser beams. The resulting
time-dependent optical potential is
\begin{equation}
\label{equ:potential1}
    V(z,t)= V_0 \cos[K(t)z-\delta(t)t],
\end{equation}
where
\begin{equation}
    K(t) =  k_{\rightarrow}(t) + k_{\leftarrow}(t)
\end{equation}
is the sum of the instantaneous wave numbers of the left- and
right-propagating laser fields forming the lattice, and
\begin{equation}
    \delta(t) = \omega_{\rightarrow}(t) - \omega_{\leftarrow}(t)
    =c \left [ k_{\rightarrow}(t) - k_{\leftarrow}(t)\right ]
\end{equation}
is the instantaneous frequency detuning between them. The
instantaneous phase velocity of the lattice fringes is therefore
\begin{equation}
    v_{{\rm lat}}(t) = \frac{\partial}{\partial t} \left
    (\frac{\delta(t) t}{K(t)} \right )
\end{equation}
which reduces to the usual phase velocity $\delta/K$ in the case
of constant detuning. Assuming that $v_{{\rm lat}}$ remains much
less than the velocity of light $c$ at all times we have
\begin{equation}
    k_{\rightarrow}(t) - k_{\leftarrow}(t) \ll
     k_{\rightarrow}(t) + k_{\leftarrow}(t),
\end{equation}
and it is an excellent approximation to set $k_{\rightarrow}(t) +
k_{\leftarrow}(t) \approx 2k_L$, where $k_L$ is a nominal laser
wave number independent of time. The potential
(\ref{equ:potential1}) reduces then to
\begin{equation}
\label{equ:potential}
    V(z,t) \approx V_0 \cos[2k_L z - \delta(t) t ]
\end{equation}
and
\begin{equation}
 v_{{\rm lat}}(t) \approx \left ( \frac{1}{2k_L} \right )
 \frac{\partial [ \delta(t) t ]}{\partial t}
\end{equation}
where the time derivative is the instantaneous frequency of the
field. While we have shown numerically that this is not
necessarily an optimum choice, we restrict our analysis in this
paper to linear accelerations produced by a time-dependent
detuning of the form
\begin{equation}
\label{equ:acc}
    \delta(t) =  \eta  t,
\end{equation}
producing a lattice group velocity
\begin{equation}
    v_{{\rm lat}}(t) =  \eta t/k_L
\end{equation}
and a constant acceleration $a_{{\rm lat}}= \eta/k_L$. This simple
case is sufficient to identify and discuss the major physical
mechanisms at play in the acceleration of the atoms.

To determine the dynamics of an ensemble of classical point
particles on this lattice, we have numerically solved Newton's
equations of motion for a large number of atoms initially at rest
at random points in an interval large compared to the lattice
period. The size of this interval and the probability of finding an
atom at a given point were chosen to mimic the shape of a Bose-Einstein
condensate density profile, to which we turn shortly.

\begin{figure}
\begin{center}
\includegraphics[width=0.8\columnwidth]{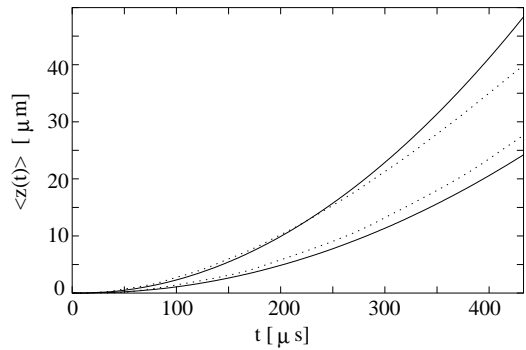}
\vspace{.3 cm} \caption{ Mean displacement of the atomic cloud:
The simulation uses $10^5$ Sodium atoms initially at rest with a
Gaussian spatial distribution of longitudinal width $50 \mbox{
}\mu\mbox{m}$ and transverse width $5 \mbox{ }\mu\mbox{m}$. The
laser wavelength is $\lambda_L = 985 \mbox{ nm}$ and the lattice
depth $V_0$ is chosen to be half the lattice recoil energy. 
The Sodium mass is $m=3.82\cdot 10^{-26} \mbox{ kg}$ and the s--wave 
scattering length $a_s=4.9 \mbox{ nm}$. Two upper
curves: $a_{{\rm lat}}= 724.1\mbox{ ms}^{-2}$. Two lower curves:
$a_{\rm {lat}}= 327.8\mbox{ ms}^{-2}$. Solid lines: classical
particles, dotted lines: BEC. Unless otherwise stated, the same
parameters are used through Fig. 7. \label{fig:zt} }
\end{center}
\end{figure}

\begin{figure}
\begin{center}
\includegraphics[width=0.8\columnwidth]{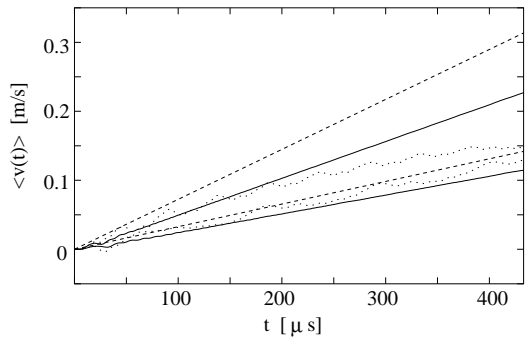}
\vspace{.3 cm}
\caption{ Mean velocity of the atomic cloud: $a_{{\rm lat}}=
  724.1\mbox{ ms}^{-2}$ (three upper curves), $a_{\rm {lat}}=
  327.8\mbox{ ms}^{-2}$ (three lower curves). Solid lines: classical
particles; dotted lines: BEC; dashed lines: instantaneous
lattice velocity. \label{fig:vt} }
\end{center}
\end{figure}

\begin{figure}
\begin{center}
\includegraphics[width=0.8\columnwidth]{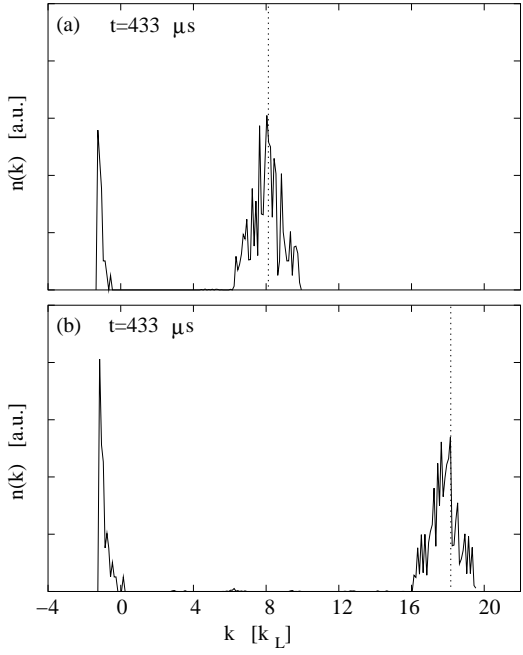}
\vspace{.3 cm} \caption{Momentum distribution $n(k)$ of an ensemble of
$10^5$ classical particles for $a_{{\rm lat}}$=(a) $327.8\mbox{
ms}^{-2}$, (b) $724.1\mbox{ ms}^{-2}$, in units of $k_L$. Vertical
dotted line: instantaneous lattice momentum at time of plot.
\label{fig:modesclassical} }
\end{center}
\end{figure}

The solid lines in Figs. \ref{fig:zt} and \ref{fig:vt} show the
evolution of the mean position $\langle z(t)\rangle$ and mean
velocity $\langle v(t)\rangle$ of the classical atomic ensemble
over the acceleration time for two values of $a_{{\rm lat}}$. We
observe that the mean atomic velocity is always somewhat less than
the instantaneous lattice velocity $v_{{\rm lat}}$, the dashed
line in Fig. \ref{fig:vt}. The cause of this difference is
revealed in Fig. \ref{fig:modesclassical}, which shows the
momentum distribution of the classical ensemble $n(k)$ at a fixed
time $t$ and for the same two values of $a_{{\rm lat}}$, where we
expressed the momentum $p$ in terms of the wavenumber $k=p/\hbar$.
In addition to peaks at positive momenta indicative of accelerated
atoms that contribute to an increase in the mean velocity of the
sample, a significant group of particles acquire negative momenta,
i.e. they are accelerated in the direction opposite to the lattice
motion. These are atoms that spill into a well to their left in
the moving potential. A similar effect
will also be encountered in the synchronous particle accelerator
of section \ref{sec:Sync}.

Another important feature of Fig. \ref{fig:modesclassical} is that
the momentum distributions of both the accelerated and decelerated
groups of atoms are rather wide. This shows that (even ignoring
the decelerated atoms), the lattice accelerator produces a
considerable heating of the atomic sample. This can potentially be
a very serious problem for applications requiring a high degree of
spatial coherence of the atomic beam. Surprisingly perhaps, we
will see that the situation can be significantly improved in the
quantum regime, a result of quantum interferences.

%-------------------------------------------
\subsection{Condensate acceleration}
\label{subsec:GPE}
%-------------------------------------------

We now turn to the acceleration of a Bose--Einstein condensate at
temperature $T=0$. For the present study, it is sufficient to
describe its dynamics in a Hartree mean-field approximation via
the Gross-Pitaevskii equation for the normalized condensate wave
function $\Psi(z,t)$,
\begin{eqnarray}
    \label{equ:GPE}
    i\hbar \frac{\partial}{\partial t}\Psi(z,t)
    &=& \left[
-\frac{\hbar^2}{2m}\frac{\partial^2}{\partial z^2} +V_0
\cos[2k_Lz-\delta(t) t] \right] \Psi(z,t) \nonumber \\ &+& N U_0
\left|\Psi\right|^2 \Psi(z,t),
\end{eqnarray}
where $N$ is the number of atoms in the condensate. For the
reduction to one dimension we assume the condensate to be in its
transverse ground state $u_g({\bf r}_\perp)$. This is accounted
for by the effective nonlinear parameter $U_0$,
\begin{equation}
\label{equ:effU0}
    U_0 = \frac{4\pi \hbar^2 a_s}{m} \int d{\bf r}_\perp \left|u_g({\bf
r}_\perp)\right|^4,
\end{equation}
where $a_s$ is the $s$-wave scattering length.

In the presence of a periodic potential, one needs to beware of
the possible fragmentation of the condensate via a Mott insulator
transition. This problem has been investigated in Ref. \cite{Jaks98},
which discusses the transition to this phase from the superfluid
phase as a function of the ratio $\kappa$ between the depth of the
optical lattice and the inter-well tunneling matrix element. The
first Mott transition, corresponding to an occupation of one atom
per lattice site, occurs for $\kappa \approx 5.8 z$, $z$ being the
number of nearest neighbors \cite{Jaks98}, corresponding typically
to lattice potential depths of the order of 2.5-5 lattice recoil
energies. Here we consider only situations where the lattice
potential is shallow enough that such fragmentation does not occur.

We solve Eq.~(\ref{equ:GPE}) numerically using a split-operator
technique for a condensate of the same initial density profile as
the classical ensemble considered earlier. The resulting mean
position $\langle z(t)\rangle$ and mean velocity $\langle
v(t)\rangle$ of the condensate are shown by the dotted lines in
Figs. \ref{fig:zt} and \ref{fig:vt}. We observe that whereas the
displacement dynamics is very similar to the classical case, the
mean velocity exhibits two differences. First, the condensate
shows pronounced time-dependent oscillations, the well--known Bloch
oscillations. Second, the higher
of the two lattice accelerations causes $\langle v\rangle$ to 
saturate to a constant value, an effect due to increased Landau-Zener
tunneling.\footnote{Superfluid effects are not an issue in
our system: The critical velocity for typical Bose--Einstein
condensates was recently determined to lie in the $\mbox{mm/s}$
regime \cite{Rama99,Onof00,Burg01}, much less than the velocities
to which the condensate is accelerated.  For the acceleration
rates under consideration here, the instantaneous lattice velocity
exceeds the critical velocity already after a small fraction of
the total acceleration time.}

The difference between the classical and quantum situations is
evidenced even more strikingly in Fig. \ref{fig:modesquantum}.
Instead of being composed of two broad continua as in Fig.
\ref{fig:modesclassical}, the condensate momentum distribution
$\phi(k)$, which is the Fourier transform of the spatial
wavefunction $\Psi(z)$, consists of a series of very narrow peaks
located at integer multiples of $2k_L$.\footnote{The experiment of
Ref. \cite{Mors01} shows clear evidence of such peaks.} Physically,
this is due to the fact that while the classical particles probe
the local value of the lattice potential, the ultracold
Bose-Einstein condensate probes a region of the lattice comprising
a large number of maxima and minima, and hence its full
periodicity. As such, the individual classical particles are
channeled in one or the other potential well of the time-dependent
potential, while the ultracold atoms are diffracted by it. Their
dynamics is governed by the quantum interferences that give rise
to Bragg scattering, with the resulting narrow peaks of Fig.
\ref{fig:modesquantum}. This is a central result of this paper: it
shows that after acceleration, the ultracold atoms can still be
largely monochromatic. The next section takes advantage of the
physical difference in physics between the acceleration of quantum
and semiclassical samples to investigate ways to achieve this
goal.

%-------------------------------------------
\subsection{Sequential Bragg resonances}
\label{subsec:Bragg}
%-------------------------------------------

\begin{figure}
\begin{center}
\includegraphics[width=0.8\columnwidth]{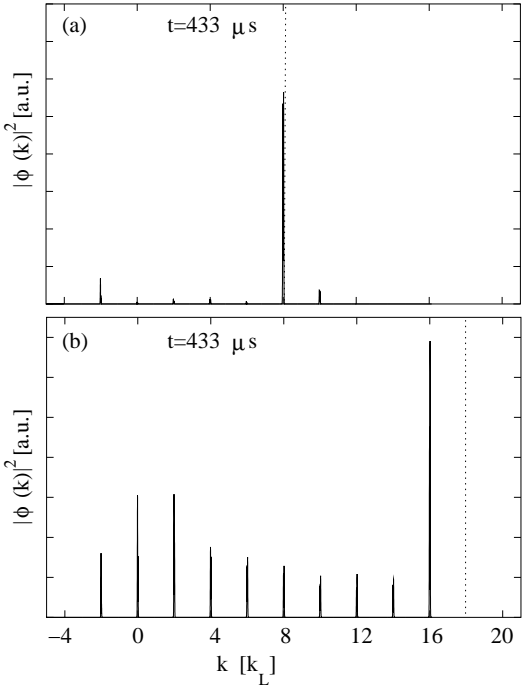}
\vspace{.3 cm}
\caption{ Momentum distribution $\phi(k)$ of a Bose--Einstein
condensate of $10^5$ atoms for $a_{{\rm lat}}$=(a) $327.8\mbox{
ms}^{-2}$, (b) $724.1\mbox{ ms}^{-2}$, in units of $k_L$. Vertical
dotted line: instantaneous lattice momentum at time of plot.
\label{fig:modesquantum} }
\end{center}
\end{figure}

With atom interferometric applications in mind, we call an
acceleration scheme ideal if it leaves all atomic population in
one single momentum mode whose value is determined by the velocity
$v_{{\rm lat}}$ of the optical lattice. We can therefore define a
figure of merit of the accelerator as one minus the fraction of
atoms in other momentum sidemodes. Fig. \ref{fig:modesquantum}
suggests that we have to restrict ourselves to certain
acceleration rates and times in order not to lose too many atoms.
The closest we were able to approach ``perfect'' acceleration with
our simple linear acceleration scheme is the example of Fig.
\ref{fig:modesquantum}(a).

In the following we consider a simple coupled-mode description of
the acceleration process to identify the important parameters in
its optimization. We neglect the effects of the mean-field energy
for now and extend the case of a static lattice as shown in
Ref. \cite{Blak00} to accelerated ones. Starting from the
collisionless form of the Gross-Pitaevskii
equation (\ref{equ:GPE}) and introducing the momentum space
condensate wave function that we already used in Fig. \ref{fig:modesquantum}
\begin{equation}
\label{equ:fourier}
\phi(k,t)
=
\frac{1}{\sqrt{2\pi}}\int\limits_{-\infty}^{\infty} {d}z\,
\psi(z,t) e^{-ikz},
\end{equation}
one finds readily that its evolution is governed by the coupled
difference-differential equations
\begin{eqnarray}
\label{equ:kspaceGPE}
i\hbar \frac{\partial}{\partial t} \phi(k)
&=&
\frac{\hbar^2 k^2}{2m}\phi(k) \nonumber \\
&+&
\frac{V_0}{2}
\left[
\phi(k-2k_L)e^{-i\eta t^2} + \phi(k+2k_L)e^{i\eta t^2}
\right].
\end{eqnarray}
>From these equations, it is clear that states with momentum $k$
only couple to neighboring states with $k=\pm2k_L$. Thus we can
partition the momentum space into intervals of width $2k_L$ and
substitute for $\phi(k,t)$ a set of wave functions
restricted to these intervals,
\begin{equation}
\label{equ:partition}
\phi_n(k)
=
\phi(k)
\quad  \mbox{for} \quad (n-1)k_L < k \le (n+1)k_L.
\end{equation}
The partitioned wave functions $\phi_n$ are then expressed in
terms of the shifted momentum $q=k-2nk_L$ and multiplied with an
interval--dependent phase factor,
\begin{equation}
\label{equ:partitionphase}
\phi_n(q)
=
\phi_n(k)e^{-in\eta t^2}.
\end{equation}
Finally, we use the numerically established property that each of
these partial wave functions suffers a momentum spread small
compared to the mode spacing , see Fig \ref{fig:modesquantum}, to
approximate them by plane waves at $q=0$,
\begin{equation}
\label{equ:coeffs}
    \phi_n(q=0, t)= c_n(t).
\end{equation}
In units of the lattice recoil frequency $\omega_{{\rm rec}}= 2 \hbar
k_L^2/m$, we introduce the dimensionless variables
\begin{equation}
\tau = t \omega_{{\rm rec}}, \quad \tilde{\eta} =
\eta/\omega_{{\rm rec}}^2, \quad \tilde{V_0} =
V_0/\hbar\omega_{{\rm rec}},
\end{equation}
in terms of which the equation of motion Eq. (\ref{equ:kspaceGPE})
can be rewritten in matrix form as
\begin{equation}
\label{equ:system}
i\frac{\partial}{\partial \tau} {\mathbf c}(\tau)
=
V(\tau){\mathbf c}(\tau),
\end{equation}
where we defined the vector of coefficients --- which
essentially gives the probability amplitudes for the various
momentum sidemodes --- as
\begin{equation}
\label{equ:vector}
{\mathbf c}(\tau)
=
\left(
\cdots , c_{-1} , c_0 , c_{1} , \cdots
\right)^T
\end{equation}
and the coupling matrix $V(\tau)$ is given by
\begin{eqnarray}
\label{equ:matrix}
{\mathbf V}(\tau)
& = &
\left(
\begin{array}{ccccc}
\displaystyle{\ddots} & \displaystyle{\ddots} & \displaystyle{0} &
\displaystyle{} & \displaystyle{} \\
\displaystyle{\ddots} & \displaystyle{\omega_{-1}(\tau)} &
\displaystyle{\frac{\tilde{V_0}}{2}} & \displaystyle{0} & \displaystyle{} \\
\displaystyle{0} & \displaystyle{\frac{\tilde{V_0}}{2}} &
\displaystyle{\omega_{0}(\tau)} & \displaystyle{\frac{\tilde{V_0}}{2}} &
\displaystyle{0} \\
\displaystyle{} & \displaystyle{0} & \displaystyle{\frac{\tilde{V_0}}{2}} &
\displaystyle{\omega_{1}(\tau)} & \displaystyle{\ddots} \\
\displaystyle{} & \displaystyle{} & \displaystyle{0} & \displaystyle{\ddots}
& \displaystyle{\ddots}
\end{array}
\right).
\end{eqnarray}
The diagonal elements of $V(\tau)$ are responsible for the
time--dependence of the coupling matrix and are given by
\begin{equation}
\label{equ:diagonal}
\omega_n(\tau)
=
n^2-2n\tilde{\eta} \tau.
\end{equation}
Finally, we can define the detuning $\Delta_n$ between adjacent
modes as
\begin{equation}
\label{equ:detuning}
\Delta_n(\tau)
=
\omega_n(\tau)-\omega_{n-1}(\tau)
=
2n-2\tilde{\eta}\tau-1.
\end{equation}

Equation (\ref{equ:system}) describes Bragg scattering of atoms
off the periodic optical lattice. Bragg resonances occur whenever
one of the detunings $\Delta_n(\tau)$ becomes equal to zero. From
Eq. (\ref{equ:detuning}), we observe that these detunings depend
both on the scattering order $n$ and on time, this latter
dependence resulting from the acceleration of the optical
potential. As a result of the linear acceleration, neighboring
pairs of modes are therefore successively moved in and out of
resonance, so that in contrast to the case of classical particles,
the physical process underlying the atomic acceleration are
successively tuned and detuned Bragg resonances.\footnote{Ref.
\cite{Peik97b} refers to this process as `` multiple adiabatic
rapid passage,'' an equally appropriate terminology.}

\begin{figure}
\begin{center}
\includegraphics[width=0.8\columnwidth]{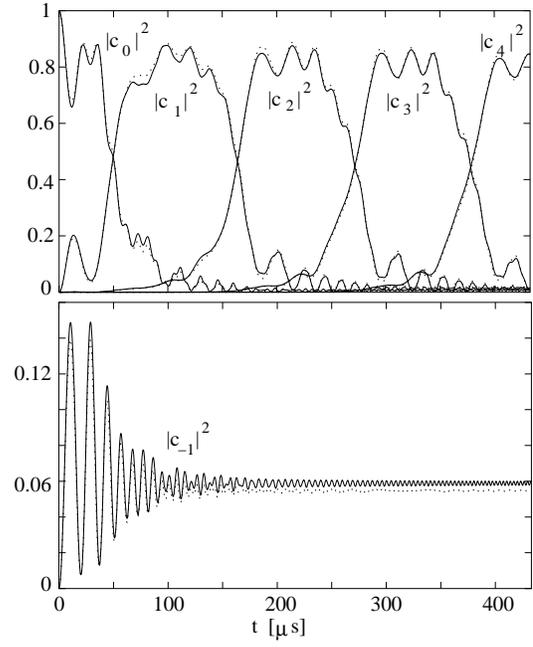}
\vspace{.3 cm} \caption{ Temporal evolution of the mode population
for an acceleration of $327.8\mbox{ ms}^{-2}$. Upper graph: $n=0$
to $n=4$ modes. Lower graph: $n=-1$ mode. Solid line: solution of
truncated coupled-mode equations; dotted line: direct solution of
the Gross-Pitaevskii equation. \label{fig:modedynamics} }
\end{center}
\end{figure}

This sequence of resonances is illustrated in Fig.
\ref{fig:modedynamics}, which shows the evolution of the
population dynamics of a few momentum sidemodes of the condensate.
The solid lines give the results of a truncated coupled-modes
analysis, while the dotted curves show the results of the direct
solution of the Gross-Pitaevskii equation, in which case the
various sidemode populations are calculated from
\begin{equation}
\left|c_n(\tau)\right|^2
=
\int\limits_{(2n-1)k_L}^{(2n+1)k_L} {d}k\,
\left|\phi(k,\tau)\right|^2.
\end{equation}
The two approaches are in excellent agreement, despite
the fact that the coupled-mode analysis included only eight modes
in the present example. The upper graph clearly illustrates the
sequential population transfer towards sidemodes of higher
momentum. The lower graph focuses on the first side mode of
negative momentum. After initial oscillations, indicative of its
coupling to other modes, its population stabilizes to a final
value resulting from the fact that it is then far--off resonance
from any other mode.

%-------------------------------------------
\subsection{Optimal acceleration}
\label{subsec:Opt}
%-------------------------------------------

Having identified sequential Bragg resonances as the physical
mechanism of acceleration of ultracold atoms, we can now attempt
to optimize the external parameters to achieve maximum velocity
and  figure of merit of the accelerator. Ignoring for now all but
the two adjacent modes $c_n$ and $c_{n-1}$, we observe that they
couple with a time-dependent Rabi frequency $\Omega_n$ given by
\begin{equation}
\label{equ:rabi}
\Omega_n(\tau)
=\frac{1}{4}\sqrt{\tilde{V_0}^2+\Delta_n^2(\tau)}.
\end{equation}
We can gain some insight into the time scale of this coupling
mechanism by introducing an averaged Rabi frequency $\bar{\Omega}$
over the interval $\tau_R = 1/\tilde{\eta}$, the time it takes the
system to move from one Bragg resonance to the next, see Eq.
(\ref{equ:detuning}),
\begin{equation}
\label{equ:averagerabi}
    \bar{\Omega}= \frac{1}{\tau_R} \int
    \limits_{\tau_n}^{\tau_n+\tau_R}d \tau\, \Omega_n(\tau),
\end{equation}
where $\tau_n=(n-1)/\tilde{\eta}$. Note that $\bar{\Omega}$ is independent
of the index $n$. Hence, one can expect that a close to optimal
mode-to-mode coupling should correspond to
\begin{equation}
\label{equ:opttau} \bar{\Omega}\tau_R \approx \pi ,
\end{equation}
since in this case the system can complete half of an averaged
Rabi cycle in the time it takes to move from one Bragg resonance
to the next. The complete population of one mode is then
approximately transferred to the next mode, thereby increasing the
momentum of the condensate by  $2\hbar k_L$. If the lattice
acceleration is too fast for this Rabi transfer to fully occur,
lower momentum modes remain significantly populated. This is the
case in the example of Fig. \ref{fig:modesquantum}(b). Conversely,
atoms accelerated too slowly undergo more then one half Rabi cycle
between two modes. This results in oscillations in the expectation
values $\langle v \rangle$ and $\langle z\rangle$.

We have mentioned in the context of Fig. \ref{fig:modedynamics}(b)
that once a momentum sidemode is shifted out of resonance, its
population remains constant. This accounts for the saturation in
$\langle v \rangle$ shown in Fig. \ref{fig:vt}. In contrast to the
classical case where all atoms
that are captured in a potential well remain captured, here, in the
quantum regime, atoms initially accelerated may gradually
be moved out of resonance, after which they retain a constant
velocity.

\begin{figure}
\begin{center}
\includegraphics[width=0.8\columnwidth]{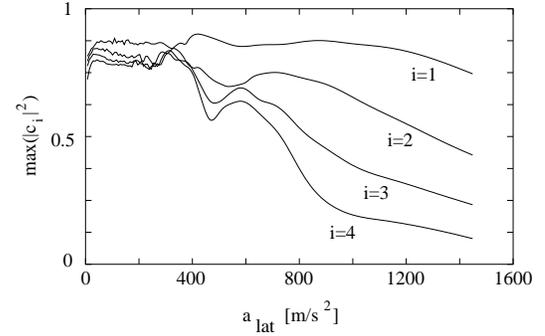}
\vspace{.3 cm} \caption{ Maximum population of the $i$th mode as a
function of acceleration, obtained from the truncated coupled-mode
equations. \label{fig:maxmode} }
\end{center}
\end{figure}

Fig. \ref{fig:maxmode} summarizes the results of a numerical
optimization of the lattice acceleration rate. It confirms that if
the acceleration rate is too large one can only transfer a small
fraction of the population to the next mode, while if it is too
small we couple to modes with negative momentum, as manifested in
the small dip in all curves near the origin. The plateau-like
feature defines the regime of efficient coupling, where around
85\% of the population is transferred.

We mention that it is also possible to accelerate the condensate
by applying a series of Bragg pulses, each of them resonant with
the next higher pair of momentum modes. For the specific
parameters used in our figures, one could then reach accelerations
approximately twice as high as in the case of continuous detuning,
but further improvement would require stronger Bragg pulses that
create additional couplings to different modes \cite{Blak00}. We
do not go into further details about this approach here. Rather,
future work will implement learning algorithms to determine ideal
chirping and/or pulse sequences to achieve the coherent and
lossless acceleration of ultracold atomic samples.

\begin{figure}
\begin{center}
\includegraphics[width=0.8\columnwidth]{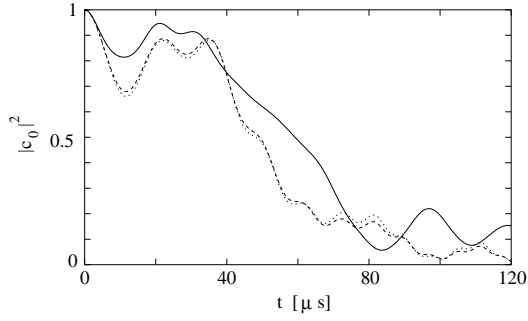}
\vspace{.3 cm} \caption{ Population of the zeroth momentum mode for
$10^6$ (solid), $10^5$ (dotted) and $10^4$ atoms (dashed) in the
condensate, obtained from a simulation of the full Gross-Pitaevskii
equation for an acceleration $a_{{\rm lat}}$=(a) $327.8\mbox{
ms}^{-2}$. \label{fig:modenonlinear}}
\end{center}
\end{figure}

We conclude this section by remarking that so far, we have ignored
the effects of the mean-field energy on the Bragg acceleration of
the atomic sample. But it should be expected that the associated
nonlinear phase shifts can effect it significantly. This is
illustrated in Fig. \ref{fig:modenonlinear},
where we plot the temporal evolution of the
zeroth momentum sidemode for different mean-field energies (atomic
numbers). For large mean-field energies we observe that the local
evolution can deviate significantly from the low-density limit,
although the general behavior is similar. For the Sodium parameter
used in our simulations, we find that in condensates of up to
$10^5$ atoms the nonlinear effects are small enough that the
system obeys the predictions of the simplified linear model.

%===========================================
\section{Synchronous particle acceleration}
\label{sec:Sync}
%===========================================
We now turn to our second accelerator scheme, which uses a
time-varying conservative potential well, formed, e.g., by
external magnetic fields. The idea is to initially place the
condensate near the top of a potential well. As it moves down the
well, it gains kinetic energy at the expense of its potential
energy. If the external field is turned off precisely when the
atoms are near the potential minimum, then the condensate will
wind up with an increased kinetic energy. The process can be
repeated by letting the condensate pass through a sequence of
pulsed potential wells appropriately displaced in space and time.
It can thus gain a considerable amount of kinetic energy.

\begin{figure}
\begin{center}
\includegraphics[width=0.8\columnwidth]{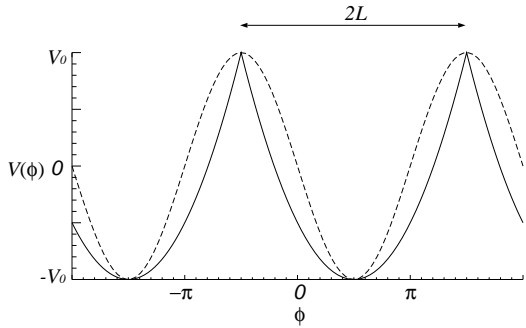}
\vspace{.3 cm}
\caption{
Potential energy as a function of phase angle. The dashed line represents a
sinusoidal potential, the solid line represents a chained harmonic
potential. For
both cases, the well depth is $2V_0$. \label{potential} }
\end{center}
\end{figure}

We first discuss how such a method can be used to accelerate an
ensemble of classical particles. Fig.~\ref{potential} illustrates
the potential energy, $V(z)$, as a function of particle position.
In analogy to the concepts of charged particle accelerators, we
express the atomic position in terms of a ``phase angle'' $\phi$,
with a periodicity of $2L$. Two examples of potentials are
depicted in Fig.~\ref{potential}: a sinusoidal potential and a
chained harmonic one. The potential maxima occur at
$\phi=-\pi/2+2n \pi$ ($n=0,\pm 1,\pm 2 ...$), while the minima are
at $\phi=\pi/2+2n \pi$. Now suppose a particle initially located
near the potential maximum accelerates to the potential minimum
for a time $\tau$, at which moment the original potential is
switched off and a new set of potential wells identical to the
previous ones, but with a phase shift of $\pi$, is switched on.
Then the particle will find itself again near the top of the
potential well, and continue to be accelerated.

In order to describe quantitatively this acceleration scheme, it
is convenient to introduce the concept of the {\em synchronous
particle}. This is a particle that travels exactly the distance
$L$ (or $\pi$ in terms of phase angle) during the time interval
$\tau$. We denote its phase and velocity right after the switching
by $\phi_s$ and $v_s$, respectively. By definition, then, the
initial phase $\phi_s$of the synchronous particle remains
unchanged.

\begin{figure}
\begin{center}
\includegraphics[width=0.8\columnwidth]{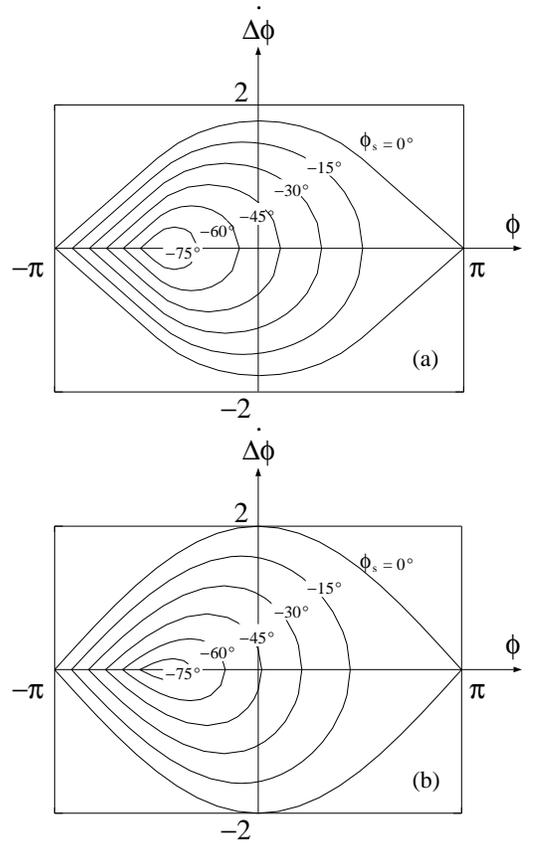}
\vspace{.3 cm}
\caption{
Phase stability diagram for various values of $\phi_s$ for (a) the harmonic
potential and (b) the sinusoidal potential. The unit for $\dot{\Delta \phi}$
is
$\sqrt {2\pi V_0/(mL^2)}$.\label{diagram} }
\end{center}
\end{figure}

The kinetic energy gained by the synchronous particle during one
acceleration stage is given by
\begin{equation}
     \Delta E_K (\phi_s) \equiv V(\phi_s) - V(\phi_s+\pi).
\end{equation}
We can regard this increase as originating from a continuously
acting, average force\cite{Beth00}
\[ F(\phi_s) = \Delta E_K(\phi_s)/L .\]
Consider now a non-synchronous particle with initial phase $\phi=
\Delta \phi + \phi_s$ and velocity $v=\Delta v+v_s$. Its motion
relative to the synchronous particle is governed approximately by
the equation of motion
\begin{equation}
\frac{d^2}{d^2t}\Delta \phi = \frac{\pi}{mL}[F(\phi_s+\Delta
\phi)-F(\phi_s)].
\label{synmotion}
\end{equation}
Depending on the initial conditions, Eq.~(\ref{synmotion})
describes either stable or unstable dynamics of the motion of a
collection of particles. It is easy to see that for stable
acceleration, we must have
\[ -\pi/2 \le \phi_s \le 0 .\]
Fig.~\ref{diagram} plots the phase stability diagram for the two
potentials depicted in Fig.~\ref{potential}, obtained by
numerically solving Eq.~(\ref{synmotion}). For a given $\phi_s$,
the closed curve depicts the boundary of a ``bucket'', inside of
which a stable acceleration is achieved. Since $\Delta E_K
(\phi_s)$ decreases from $2V_0$ to 0 when $\phi_s$ changes from
$-\pi/2$ to 0, we see there is always a trade-off between fast
acceleration and large bucket size, i.e., large acceptance.
Furthermore, for $-\pi/2 < \phi_s < -\pi/12$, the chained harmonic
potential gives a larger bucket compared to the sinusoidal
potential.

\begin{figure}
\begin{center}
\includegraphics[width=0.8\columnwidth]{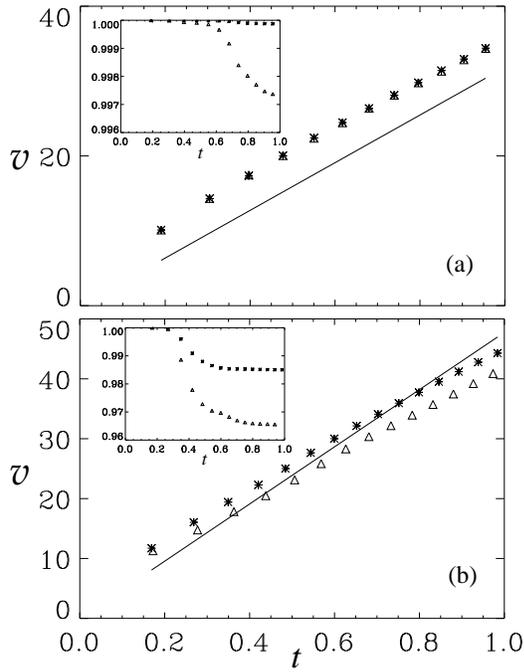}
\vspace{.3 cm} \caption{ Condensate velocity at the times of the
successive switchings for (a) the harmonic potential and (b) the
sinusoidal potential. Stars: non-interacting condensate with
$U=0$; triangles: interacting condensate with $U=10$. For
practical choices of parameters, the typical value for $U$ in our
dimensionless units is on the order of 1 to 10. Inserts:
normalization of the wavepackets. Time in units of $2mL^2/(\hbar
\pi^2)$, velocity in units of $\hbar \pi/(mL)$, and energies in
units of $\hbar^2\pi^2/(2mL^2)$.  In this example, $V_0=150$. (In
practice, $V_0$ should have a much larger value for realistic
choices of parameters. However, in our calculations large values
of $V_0$ gives rise to too much numerical noise.)\label{velocity}
}
\end{center}
\end{figure}

To test our scheme on a condensate, we have solved the
time-dependent Gross-Pitaevskii equation (\ref{equ:GPE}) with the
time-dependent potential of Fig.~\ref{potential}. Our results are
illustrated in Fig.~\ref{velocity}, which shows the velocity of
the condensate at each of the switching times of the external
fields, for both harmonic and sinusoidal potentials. The switching
is chosen to occur at the moment the center of mass of the
condensate travels a distance $L$. For this calculation, we choose
the initial wave function to be a Gaussian centered at
$\phi_0=-\pi/6$ with a width of $0.1L$. The inserts in the figure
show the normalization of the wave function, which characterizes
the loss of atoms during their acceleration. This is calculated
by discarding the atoms leaked
outside of the well that traps the center of the mass of the
condensate.
Less loss is observed
for the harmonic potential than for the sinusoidal one, a result
of the larger bucket size in the former case (see
Fig.~\ref{diagram}). We also observe a weaker influence of the
condensate mean-field energy in the case of a harmonic potential.
However, while atomic collisions do increase the losses and
decrease acceleration rates, their effect is not very dramatic, a
result similar to that already encountered with the chirped
lattice accelerator. In addition, we checked the momentum
distribution of the wave function during the acceleration and
found no significant variations of the momentum width, hence,
monochromaticity is also preserved in this scheme.

The solid line in Fig.~\ref{velocity} gives the velocity of a
classical particle under the action of a continuous and constant
force $F(\phi_0)=\Delta E_K(\phi_0)/L$. As the figure shows, the
average acceleration $a$ of the condensate can indeed be
approximated by $a=F(\phi_0)/m$.

We finally note that in the case of alkali atoms, an acceleration
of $\sim 10^5$ m/s$^2$ requires a well depth of 10 mK and width
$L=100$ $\mu$m is needed. In view of current advances in the
technology of magnetic microtraps\cite{Reic01}, these numbers
should not pose as an unsurmountable challenge for
experimentalists.

%===========================================
\section{Summary and conclusion}
\label{sec:SumCon}
%===========================================

In conclusion, we demonstrated two coherent mechanisms to
accelerate Bose--Einstein condensates with very little atom loss
while preserving its monochromaticity. The first mechanism, based
on the use of an accelerated optical lattice, can be interpreted
as resulting from a sequence of Bragg resonances, which can lead
to a remarkable preservation of the monochromaticity of the atomic
sample under appropriate conditions. This analysis enabled us to
identify an optimal regime of acceleration rates for reasonable
experimental parameters. The second scheme uses a time-varying
potential analogous to those used in the synchronous acceleration
of charged particle. Acceleration is achieved by switching on and
off the potential in an appropriate way such that the condensate
keeps gaining kinetic energy at the expense of its potential
energy. In contrast to the first scheme where only quasi--discrete
modes are populated, here the momentum of the condensate evolves
continuously.

In future work, it will be important to optimize the
time-dependence of the accelerators, so as to achieve a factor of
merit as close to unity as possible. We propose to approach this
problem using genetic learning algorithm techniques. It will also
be necessary to go past the mean-field theory in order to
determine the higher-order coherence properties of the accelerated
beams, in particular their atom statistics. In addition, these
acceleration devices are potentially important to generate
non-classical matter wave fields to be used in improving the
signal to noise ratio in future atom optical sensors. The discrete
set of momentum sidemodes that are macroscopically excited in the
sequential Bragg resonances make this problem particularly well
suited for quantum optical ``essential modes'' approaches familiar
from quantum nonlinear optics.

\acknowledgements

This work is supported in part by the U.S. Office of Naval
Research under Contract No. 14-91-J1205, by the National Science
Foundation under Grant No. PHY98-01099, by the U.S. Army Research
Office, and by the Joint Services Optics Program.

\newpage

\newpage

\vspace{5 cm}

\end{document}